\documentclass{article} 
\usepackage{amsmath}
\usepackage{graphicx} 
\usepackage[utf8]{inputenc}
\usepackage {comment}
\usepackage{authblk}
\usepackage{lineno}
\usepackage{color}

\newcommand{\abs}[1]{\left|#1\right|} 
\newcommand{\avg}[1]{\left\langle#1\right\rangle} 

\newcommand{\ludremoved}[1] {}

\begin{document}


\title{Stochastic oscillations produce dragon king avalanches in 
self-organized quasi-critical systems}

\author{Osame Kinouchi$^{1*}$, Ludmila Brochini$^2$,
Ariadne de Andrade Costa$^3$,
João Guilherme Ferreira Campos$^4$ \& 
Mauro Copelli$^4$}


\affil[1]{Universidade de São Paulo, 
Departamento de Física-FFCLRP, Ribeirão Preto, Brazil} 
\affil[2]{Universidade de São Paulo, Instituto de Matemática
e Estatística, São Paulo, Brazil}
\affil[3]{Universidade de Campinas, 
Instituto de Computação, Campinas, Brazil}
\affil[4]{Universidade Federal de Pernambuco, 
 Departamento de Física, Recife, Brazil}



\maketitle
\begin{abstract}
In the last decade, several models with network 
adaptive mechanisms (link deletion-creation,
dynamic synapses, dynamic gains) have been proposed as examples of
self-organized criticality (SOC) to explain neuronal 
avalanches. However, all these systems
present stochastic oscillations hovering around the critical 
region that are incompatible with standard SOC. 
This phenomenology has been called 
self-organized quasi-criticality (SOqC).
Here we make a linear stability analysis of the mean field 
fixed points of two SOqC systems: a fully connected network 
of discrete time stochastic spiking neurons with firing rate 
adaptation produced by dynamic neuronal gains and an excitable 
cellular automata with depressing synapses. We find that the 
fixed point corresponds to a stable focus that loses stability 
at criticality. We argue that when this focus is close 
to become indifferent, demographic noise can elicit stochastic 
oscillations that frequently fall into
the absorbing state. This mechanism interrupts the oscillations, 
producing both power law avalanches and dragon king events, which
appear as bands of synchronized firings in raster plots.
Our approach differs from standard SOC models in that it 
predicts the coexistence of these different 
types of neuronal activity.

\end{abstract}

\section{Introduction}

Conservative self-organized critical (SOC) systems are by now
well understood in the framework of out of equilibrium
absorbing phase transitions~\cite{Dickman1998,Jensen1998,
Dickman2000,Pruessner2012}.
But, since natural systems that 
present power law avalanches (earthquakes,
forest fires etc.) are dissipative in the bulk,
compensation (drive) mechanisms have been proposed to make the 
models at least conservative on average. 
However, it seems that these mechanisms do not work so well:
they produce ``dirty criticality", 
in which the system hovers around the critical region with stochastic 
oscillations (SO)~\cite{Bonachela2009,Bonachela2010}.

The literature reserves the acronym aSOC (adaptive SOC)
to models that have an explicit 
dynamics in topology or network parameters
(link deletion and creation, adaptive synapses, 
adaptive gains)~\cite{Bornholdt2000,Meisel2009,Meisel2012,Droste2013, Brochini2016,Costa2017}. 
In the area of neuronal 
avalanches~\cite{Beggs2003,Chialvo2004,Beggs2008,Chialvo2010}, 
a well known aSOC system that presents SO is the
Levina, Herrmann and Geisel (LHG)
model~\cite{Levina2007,Levina2009,Bonachela2010} 
which uses continuous time integrate-and-fire neurons and 
dynamic synapses with short-term depression 
inspired by the work of Tsodyks and 
Markram~\cite{Tsodyks1997,Tsodyks2006}.
After that, similar models have been studied, for example
excitable cellular automata~\cite{Kinouchi2006} 
with LHG synapses~\cite{Costa2015,Campos2017} 
and  discrete time stochastic neurons  
with dynamic neuronal gains~\cite{Brochini2016,Costa2017,Costa2018}.

Not all SOqC systems (say, forest-fire models) are aSOC 
(which always have adaptive network parameters)
but it seems that all aSOC systems 
are examples of SOqC.
The exact origin of SO in aSOC systems
is a bit unclear~\cite{Bonachela2009,Bonachela2010,
Costa2017,Costa2018}.
Here, we examine some representative discrete time aSOC 
models at the mean-field (MF) level.
We find that they evolve as 2d MF maps whose
fixed point is a stable focus very close to 
a Neimark-Sacker-like bifurcation, which defines the critical point. 

This kind of stochastic oscillation is known in the 
literature, sometimes called quasicycles~\cite{Nisbet1976,McKane2005,Risau2007,
Wallace2011,Baxendale2011,Challenger2014,Parra2014} 
but, to produce them, one ordinarily needs to fine tune the 
system close to the bifurcation point. 
In contrast, for aSOC systems, there is a 
self-organization dynamics that tunes the system very close to the critical 
point~\cite{Levina2007,Bonachela2010,Costa2015,
Campos2017,Brochini2016,Costa2017}. 

Although aSOC models show no 
exact criticality, they are very interesting because they can
explain the coexistence of power law distributed avalanches, 
very large events (``dragon
kings"~\cite{Arcangelis2012,Orlandi2013,Yaghoubi2018})
and stochastic oscillations~\cite{Costa2017,Costa2018}.
Also, the adaptive mechanisms are biologically plausible and
local, that is, they do not use non-local information to
tune the system toward the critical region as occurs in other
models~\cite{Arcangelis2006,Arcangelis2012b}.

\section{Network model with stochastic neurons}

Our basic elements are discrete time stochastic integrate-and-fire
neurons~\cite{Gerstner1991,Gerstner1992,Gerstner2002,
Galves2013,Larremore2014}. They enable
simple and transparent analytic results~\cite{Brochini2016,Costa2017} 
but have not been intensively studied.  
We consider a fully connected topology with
$i = 1, \ldots,N$ neurons.
Let $X_i$ be a firing indicator:  $X_i[t]=1$ means 
that neuron $i$ spiked at time $t$ and $X_i[t]=0$ 
indicates that neuron $i$ was silent at time $t$.  
Each neuron spikes with a firing probability 
function $\Phi(V_i[t])$ that depends on a real valued variable $V_i[t]$ 
(the membrane potential of neuron $i$ at time $t$). 
Notice that, although the firing indicator is binary, 
the model is not a binary cellular
automaton but corresponds to a stochastic version of leaky 
integrate-and-fire neurons.

\begin{figure}[ht]
\includegraphics[width=\columnwidth]{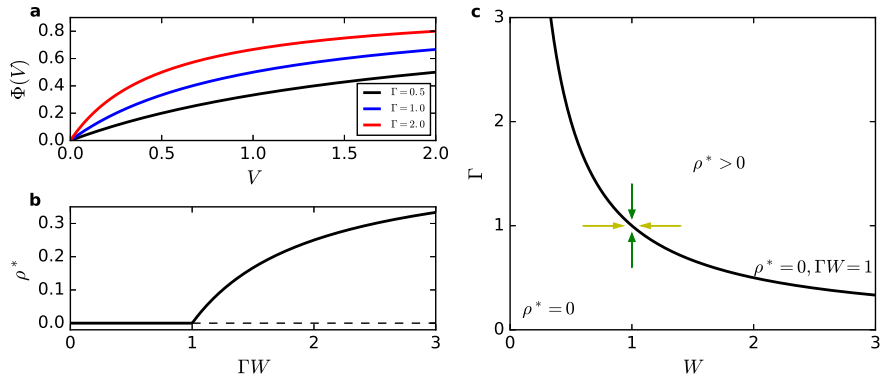}
    \caption{{
    \bf Firing function $\Phi(V)$, firing density and phase diagram
    for the static model.} 
    {\bf a}, Rational firing function
      $\Phi(V)$ for $\Gamma = 0.5$ (bottom), 
      $1.0$ (middle) and $2.0$ (top).
  {\bf b}, Firing density $\rho(\Gamma W)$. 
  The absorbing state $\rho^0=0$ looses stability after 
  $\Gamma W > \Gamma_c W_c = 1$. 
  {\bf c}, Phase diagram in the $\Gamma \times W$ plane. 
  An aSOC network can be created by adapting synapses 
  (horizontal arrows) or adapting neuronal gains
  (vertical arrows) toward the critical line. 
  }
\label{GxW}
\end{figure}

The firing function can be any general
monotonically increasing function $0 \leq \Phi(V) \leq 1 $.
For mathematical convenience, we use the so-called 
rational function~\cite{Costa2017}:

\begin{equation}
P(X_i[t]=1|V_i[t]) = \Phi(V_i[t]) = 
\frac{\Gamma V_i[t]}{1 + \Gamma V_i[t]} \:\Theta(V_i[t]) \: , \label{Phi}
\end{equation}
where $\Gamma$ is the neuronal gain (the derivative $d \Phi/dV$ for
small $V$) and $\Theta$ is the
step Heaviside function. This firing function is shown 
in Fig.~\ref{GxW}a.

In a general case the membrane voltage evolves as:
\begin{eqnarray}
V_i[t+1]  = \begin{cases} \mu V_i[t] + I_i[t] + \frac{1}{N}\sum_{j=1}^N W_{ij} X_j[t] 
&\mbox{if } X_i[t] = 0 \:, \label{VV}  \\ 
 0 &\mbox{if }  X_i[t]=1 \label{VV0} \end{cases}
\end{eqnarray}


where $\mu$ is a leakage parameter and $I_i[t]$ is an external input.
The synaptic weights $W_{ij}$ ($W_{ii}=0$) are real valued  
with average $W$ and finite variance. 
In the present case we study only excitatory neurons ($W_{ij}>0$)
but the model can be fully generalized to an excitatory-inhibitory
network with a fraction $p$ of excitatory and $q = 1-p$ inhibitory
neurons~\cite{Kinouchi2018b}. The
voltage is reset to zero after a firing event in the previous time step.
Since $\Phi(0) = 0$, this means that two consecutive firings are
not allowed (the neuron has a refractory period of one time step).

In this paper we are interested in the second order absorbing phase 
transition that occurs when the external fields
$I_i$ are zero. Also, the universality class
of the phase transition is the same for any value of $\mu$~\cite{Costa2017},
so we focus our attention to the simplest case $\mu = 0$. 
In the MF approximation,
we substitute $X_i$ by its mean value $\rho = \avg{X_i}$, which is
our order parameter (density of firing neurons or density of
active sites). Then, 
equation~(\ref{VV}) reads $V[t+1] = W \rho[t]$, where $W = \avg{W_{ij}}$.
From the definition of the firing function Eq.~(\ref{Phi}), we have:
\begin{equation}
\rho[t] =\int \Phi(V)\: p(V)[t]\: dV \:, \label{rho}\:
\end{equation}
where $p(V)[t]$ is the voltage density at time $t$.

To proceed with the stability analysis of fixed points,
it suffices to obtain the map for $\rho[t]$ close to stationarity.
After a transient where all neurons spike at least one time, 
equations~(\ref{VV}) 
lead to a voltage density that has two Dirac peaks, 
$p(V)[t+1] = \rho[t] \delta(V) 
+ (1-\rho[t]) \delta(V - W \rho[t])$. 
Inserting $p(V)$ in equation~(\ref{rho}), we finally get: 
\begin{equation}
\rho[t+1] =  \frac{\Gamma W \rho[t] (1-\rho[t])}
{1+ \Gamma W \rho[t]} \Theta(W \rho[t]) \:.
\end{equation}
The $1-\rho[t]$ fraction corresponds to the density of silent neurons in the previous time step. We call this the static model because $W$ and $\Gamma$ are fixed control parameters.

The order parameter $\rho$ has two
fixed points: the absorbing state 
$\rho^0 = 0$, which is stable (unstable) for 
$\Gamma W < 1$, ($>1$) and a non-trivial firing state:
\begin{equation}
\rho^* = \frac{\Gamma W - 1}{ 2 \Gamma W} \:, \label{rhocrit}
\end{equation}
which is stable for $\Gamma W > 1$.
This means that $\Gamma W = 1$ is a 
critical line of a continuous
absorbing state transition (transcritical bifurcation), see 
Fig.~\ref{GxW}b and \ref{GxW}c.
If parameters are put at this critical line one observes
well behaved power laws for avalanche sizes and 
durations with the mean-field exponents
$-3/2$ and $2$ respectively~\cite{Brochini2016}. 
However, such fine tuning should not be
allowed for systems that are intended to be self-organized in criticality. 

\subsection{Stochastic neurons model with dynamic synapses}

Adaptive SOC models try to turn the critical point 
into an attractive fixed point of some homeostatic dynamics. 
For example, in the LHG model, the spike of the presynaptic neuron
produces depression of the synapse, which recovers within some 
large time  scale~\cite{Levina2007,Bonachela2010}.

A model with stochastic neurons and LHG 
synapses uses the same equations~(\ref{Phi} and \ref{VV}),
but now the synapses change with time~\cite{Brochini2016}:
\begin{equation} 
W_{ij}[t+1] =  W_{ij}[t] +\frac{\Delta t}{\tau} \left(A - W_{ij}[t]\right)
- u W_{ij}[t] X_j[t]\:. \label{LHG}
\end{equation}
Here, $\tau$ is the recovery time toward a baseline level $A$ and 
$0<u<1$ is the fraction of synaptic strength that is lost when 
the presynaptic neuron fires ($X_j = 1$). 
From now, we always use
$\Delta t = 1$ ms, the typical width of a spike.

\subsection{Stochastic neurons model with dynamic neuronal gains}

Instead of adapting synapses toward the critical line 
$W_c = 1/\Gamma$, we can adapt the gains toward the
critical condition $\Gamma_c = 1/W$, see Fig.~\ref{GxW}c. 
This can be modeled as individual 
dynamic neuronal gains $\Gamma_i[t]$ ($i = 1,\ldots,N$)
that decrease by a factor $u$
if the neuron fires (diminishing
the probability of subsequent firings) with a recovery 
time $1/\tau$ toward a baseline level $A$~\cite{Brochini2016}:
\begin{equation}
\Gamma_i[t+1] = \Gamma_i[t] + \frac{1}{\tau}
\left(A-\Gamma_i[t] \right) - u \Gamma[t] X_i[t] \:, \label{LHG-G}
\end{equation}
which is very similar to the LHG dynamics. Notice, however,
that here we have only $N$ equations for the 
neuronal gains instead of $N(N-1)$
equations for dynamic synapses, which allows the simulation
of much larger systems.
Also, the neuronal gain depression occurs due to the firing of 
the neuron $i$ (that is, $X_i[t] = 1$) instead of the 
firing of the presynaptic neuron $j$.
The biological location is also different: adaption of neuronal
gains (that produces firing rate adaptation) is a process that occurs 
at the axonal initial segment (AIS)~\cite{Kole2012,Pozzorini2013} 
instead of dendritic synapses.

Like the LHG model, this dynamics inconveniently has three parameters ($\tau, A$ and $u$). Recently,
we proposed a simpler dynamics with only one 
parameter~\cite{Costa2017,Costa2018}:
\begin{equation}
\Gamma_i[t+1] = \Gamma_i[t] + \frac{1}{\tau}
\Gamma_i[t] - \Gamma_i[t] X_i[t] =  \left(1+ \frac{1}{\tau}
- X_i[t] \right) \Gamma_i[t] \: . \label{gammat}
\end{equation}
Averaging over the sites, we obtain the 2d MF map:
\begin{eqnarray}
\rho[t+1] &=& \frac{\Gamma[t] W \rho[t] (1 - \rho[t])}
{1+\Gamma[t] W \rho[t]}\:,  \label{rhomap} \\
\Gamma[t+1] &=& \left(1+ \frac{1}{\tau}
- \rho[t] \right) \Gamma[t] \:. \label{G}
\end{eqnarray}

\subsection{Stability analysis for stochastic neurons with simplified neuronal gains}

This case with a single-parameter dynamics ($\tau$) for the neuronal gains has the simplest 
analytic results, so it will be presented first and with more
detail. For finite $\tau$, $\rho^0 = 0$ is no longer a solution, see
equation~(\ref{G}), and
the 2d map has a single fixed point $(\rho^*,\Gamma^*)$:
\begin{equation}
\rho^*  =  \frac{1}{\tau}\:,\:\:\:\:\:\:\:\:\:\:\:\:\:\:\:\:\:
\Gamma^* =  \frac{\Gamma_c}{1-2/\tau} \:, \label{GFP}
\end{equation}
where $\Gamma_c = 1/W $.
The relation between $\rho^*$ and $\Gamma^*$ is:
\begin{equation}
\rho^* = \frac{\Gamma^* W -1 }{2\Gamma^*W}  
= \frac{\Gamma^* - \Gamma_c}{2\Gamma^*} \:,
\end{equation}
which resembles the expression for the transcritical
phase transition in the static system,
see equation~(\ref{rhocrit}). 
Here, however, $\Gamma^*$ is no longer a parameter
to be tuned but rather a fixed point of the 2d map to which the 
system dynamically converges. 
Notice that the critical point of the static model can be approximated 
for large $\tau$, with $(\rho^*,\Gamma^*) 
\rightarrow_{\tau \rightarrow \infty} (0,\Gamma_c)$.

Performing a linear stability analysis of the fixed point (see
Supplementary Information), 
we find that it corresponds to stable focus. The modulus
of the complex eigenvalues is:
\begin{equation}
|\lambda^\pm| = \sqrt{1 - \frac{\tau+2}{\tau(\tau-1)}}\:.
\label{lambsimple}
\end{equation}
For large $\tau$ we have $|\lambda^\pm| = 1 - O(1/\tau)$, 
with a Neimark-Sacker-like
critical point occurring when $|\lambda^\pm| = 1$ where the focus turns out
indifferent. 
For example, we have
$|\lambda^\pm| \approx 0.990$ for $\tau = 100$, 
$|\lambda^\pm| \approx 0.998$ for $\tau = 500$ and 
$|\lambda^\pm| \approx 0.999$ for $\tau = 1,000$.
Since, due to biological motivations,
$\tau$ is in the interval of $100-1000$ ms, 
we see that the focus is at the border of
losing their stability. 
We call this point Neimark-Sacker-like because, in contrast to
usual Neimark-Sacker one, the other side of the bifurcation, with 
$|\lambda^\pm| > 1$, does not exist.

\subsection{Finite size fluctuations produce stochastic oscillations}

The stability of the fixed point focus is a result for the MF map that
represents an infinite system without fluctuations. However,
fluctuations are present in any finite size system 
and these fluctuations perturb
the almost indifferent focus, exciting and 
sustaining stochastic oscillations that hover around the fixed point. 

\begin{figure}[ht]
\begin{center}
\includegraphics[width=\columnwidth]{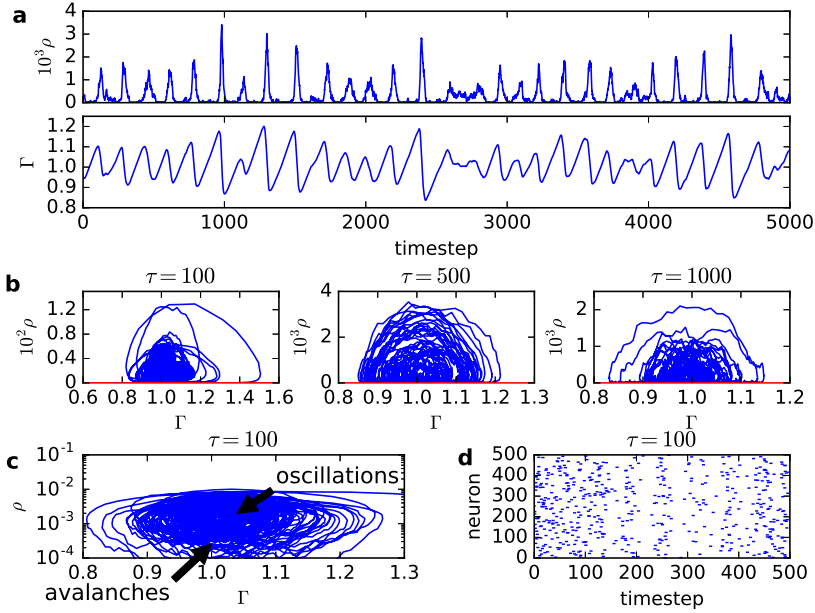}

\caption{{\bf Stochastic neurons network with simplified
(one parameter) neuronal gain dynamics (Eq.~\ref{gammat}) and $W=1$.}
{\bf a}, Stochastic oscillations for $\rho[t]$ and $\Gamma[t]$
($\tau = 500, N=100,000$ neurons). 
{\bf b}, SO in the $\rho$ vs $\Gamma$ phase plane for 
$\tau = 100, 500$ and $1,000$ ($ N=100,000$). 
{\bf c}, SO in the log $\rho$ vs $\Gamma$ phase plane for 
$\tau = 100$. {\bf d}, Raster plot with $500$ neurons for
$\tau = 100$.}
\label{HSO}
\end{center}
\end{figure}

Without loss of generality, we fix $W=1$ in the simulations
for the simple model with one parameter dynamics defined
by Eq.~(\ref{gammat}).
In Fig.~\ref{HSO}a, we show the SO for the firing
density $\rho[t]$ and average gain $\Gamma[t]$
($\tau = 500$ and $N= 100,000$). As observed in the original
LHG model~\cite{Bonachela2010},
the stochastic oscillations have a sawtooth 
profile with no constant amplitude or period. 
In Fig.~\ref{HSO}b, we show the SO in the phase plane
$\rho$ vs $\Gamma$ for $\tau = 100, 500$ and $1,000$.

For small amplitude (harmonic) oscillations, 
the frequency is given by (see Supplementary Information):
\begin{equation}
\omega  = \arctan 
\frac{\sqrt{\tau + 2/\tau - 4}}{\tau - 2} \:.
\end{equation}
The oscillation period, also for small amplitudes, is given
by $T = 2 \pi/\omega$.
The full oscillations are non-linear 
and their frequency and period depend on the amplitude.

Notice that, in the critical case, 
we have full critical slowing down: $
\lim_{\tau \rightarrow \infty} \omega =
\tau^{-1/2} \rightarrow 0,
\lim_{\tau \rightarrow \infty} T \rightarrow \infty $.
This means that, exactly at the critical point, the 2d map corresponds
to a center without a period, and the SO would correspond to
critical fluctuations similar to random walks 
in the $\rho$ vs $\Gamma$ plane.
However, since for any physical/biological system 
the recovery time $\tau$ is finite, the
critical point is not observable.

\subsection{Stochastic oscillations and avalanches}

For large $\tau$, part of the SO orbit occurs very close to
the zero state $\rho^0=0$, see Figs.~\ref{HSO}b-c. 
Due to finite-size fluctuations, the
system frequently falls into this absorbing state. 
The orbits in phase space are interrupted. 
As in usual SOC simulations,
we can define the size of an avalanche as the number 
$S$ of firing events
between such zero states. After a zero state, we force a neuron 
to fire, to continue the dynamics, so that 
Figs.~\ref{HSO}b-c are  better
understood as a series of patches (avalanches) 
terminating at $\rho[t]=0$, not as a single orbit.

\begin{figure}[ht]
\begin{center}
\includegraphics[width=\columnwidth]{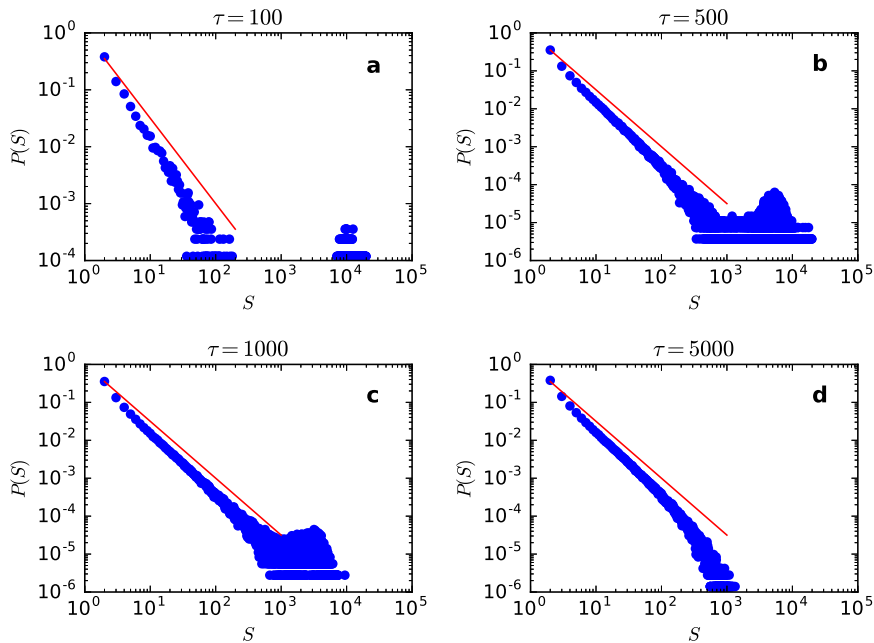}

\caption{{\bf Avalanche size distribution $P(S)$ for
the one parameter neuronal gain dynamics (Eq.~\ref{gammat}) and $W=1$.}
a) $\tau = 100$, b) $\tau = 500$, c) $\tau = 1,000$ and d)
$\tau = 5,000$; all plots
have $N=100,000$. 
The straight line corresponds to the exponent $-3/2$.
The apparent subcriticality in d) is a finite size effect.}
\label{aval}
\end{center}
\end{figure}

The presence of the SO affects the distribution of avalanche
sizes $P(S)$, see Fig.~\ref{aval}. 
For small $\tau$, $\rho^*$ is larger and
it is more difficult for the SO to fall into the absorbing state.
We observe a bump of very large avalanches (dragon king events). 
Increasing $\tau$, we move closer to the 
Neimark-Sacker-like critical point and observe power 
law avalanches with
exponent $-3/2$ similar to those
produced in the static model that suffers a transcritical bifurcation. 
So, by using a different mechanism (Neimar-Sacker-like
versus transcritical bifurcation), our 
aSOC model can reproduce experimental data about power law neuronal 
avalanches~\cite{Beggs2003,Chialvo2004,Beggs2008,Chialvo2010}.
It also predicts that these avalanches can coexist with dragon kings
events~\cite{Orlandi2013,Yaghoubi2018}.
We notice that Fig.~\ref{aval} for $\tau = 5000$ seems to be subcritical,
but this is a finite-size effect~\cite{Costa2017}.
The $P(S)$ distribution is not our main concern 
here and a more rigorous finite size analysis can be found 
in previous papers~\cite{Brochini2016,Costa2017}.

\subsection{Stochastic neurons with LHG dynamic gains}

We now return to the case of dynamic gains with LHG dynamics,
see equation~(\ref{LHG-G}). 
Without loss of generality, we use $W = 1$, so that the static model has $\Gamma_c = 1/W = 1$.
The map is given by:
\begin{eqnarray}
\rho[t+1] &=&  \frac{\Gamma[t] \rho[t] 
( 1- \rho[t])}{1+\Gamma[t] \rho[t]}\:,\\
\Gamma[t+1] & = &  \Gamma[t] 
+ \frac{1}{\tau} \left( A - \Gamma[t] \right) 
- u \Gamma[t] \rho[t]   \:.
\end{eqnarray}

The trivial absorbing fixed point is $(\rho^0,\Gamma^0) = (0,A)$,
stable for $A<1$, see Supplementary Material.
For $A > 1$, we have a non-trivial fixed point given by:
\begin{eqnarray}
\rho^*  &= & \frac{A-1}{2 A+\tau u}\:,\\
\Gamma^*  &=&  \frac{2 A+\tau u}{2+ \tau u} =  \Gamma_c 
+ \frac{2(A-1)+\tau u}{2+\tau u}\:.
\end{eqnarray}
There is a relation between $\rho^*$ and $\Gamma^*$ that
resembles the relation between the order parameter and 
the control parameter in the static model: 
\begin{equation}
\rho^*  =   \frac{\Gamma^*-1}{2 \Gamma^*} \:,
\end{equation}
valid for $\Gamma^* > \Gamma_c = 1$.
Here, however,
$\Gamma^*$ is a self-organized variable, not a control parameter
to be finely tuned.

We notice that to set by hand $A=1$, pulling all gains 
toward $\Gamma_i=1$, produces the critical
point $(\rho^*,\Gamma^*) = (0,1)$. Nevertheless, this
is a fine tune that should not be allowed for
SOC systems. We must use $ A > 1$, and reach 
the critical region only at the large 
$\tau$ limit: $ \rho^* \approx (A-1)/\tau u$ and 
$\Gamma^* \approx \Gamma_c + 2(A-1)/\tau u$.

Proceeding with the linear stability analysis, we obtain
(see Supplementary Information):
\begin{eqnarray}
\lambda^+ \lambda^- =
\left( 1 - \frac{1}{\tau} \right) 
\left(1 - \frac{2(A-1)}{A+ u \tau + 1} \right) + \frac{u (A-1)^2}
{(A + u \tau + 1)(2A + u \tau)} \:.
\label{lambda}
\end{eqnarray}

To first order in $\tau$, we have:
\begin{equation}
\abs{\lambda^\pm}=\sqrt{\lambda^+\lambda^-}=1-\frac{2(A-1)+u}{2 u\tau} 
+ O(\tau^{-2})\:, \label{lambLHG}
\end{equation}
which means that, for large $\tau$, the map is very close to the 
Neimark-Sacker-like critical point. As the stable focus approaches the critical value, the system exhibits oscillations as it approaches the fixed point due to demographic noise, leading to SO in finite-sized systems. 

\begin{figure}[ht]
\begin{center}
\includegraphics[width=\columnwidth]{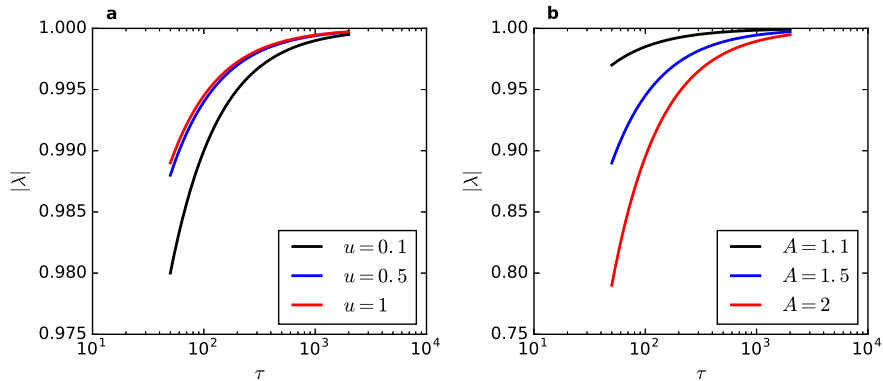}
\caption{{\bf Modulus $|\lambda^\pm|$ as a function of $\tau$
for several values of $u$ and $A$.}
{\bf a}, From top to bottom, $u = 0.1, 0.5$ and $1.0$, for $A= 1.05$.
{\bf b}, From top to bottom, $A = 1.1, 1.5$ and $A=2.0$, for $u = 0.1$.
}
\label{figlambda}
\end{center}
\end{figure}

For example, with the typical values $A=1.05$ and 
$u= 0.1$~\cite{Costa2015,Campos2017}, 
we have $\abs{\lambda^\pm} 
\approx 1 - 1/\tau$, which gives  
$\abs{\lambda^\pm} \approx 0.990$ for
$\tau = 100 $ 
and $\abs{\lambda^\pm} \approx 0.999$ for $\tau = 1,000$. 
In Fig.~\ref{figlambda}, we present the exact
$\abs{\lambda^\pm}$, the square root of 
equation~(\ref{lambda}), as a function
of $\tau$ for several values of $A$ and $u$. For large $\tau,$
the frequency for harmonic oscillations is given by
$\omega \simeq \sqrt{(A-1)/\tau} $
(see Supplementary Information).

The analysis of the stochastic neuron model 
with LHG synapses $W_{ij}[t]$, see equation~(\ref{LHG}), 
is very similar to the one above with 
LHG dynamic gains $\Gamma_i[t]$. 
The only difference is that we need to exchange $\Gamma$ by $W$. 
The eigenvalue modulus is the same, so that 
LHG dynamic synapses also produce SO. 
But, instead of presenting
simulations in our complete graph system, which would involve
$N(N-1)$ dynamic equations for the synapses, we prefer to
discuss the LHG synapses for another 
system well known in the literature.
We will also examine how SO depends on the system size $N$.

\section{Excitable cellular automata  with LHG synapses}

We now consider an aSOC version of a probabilistic 
cellular automata well studied in the 
literature~\cite{Kinouchi2006,Larremore2011,Pei2012,
Mosqueiro2013,Wang2017,Zhang2018}.
It is a model for excitable media that yields a natural interpretation in terms
of neuronal networks. Each site has $n$ states,
$X=0$ (silent), $X=1$ (firing), $X = 2,\ldots, n-1$ (refractory).
Here we will use $n=2$. 

Each neuron $i= 1,\ldots,N$ has $K_i$ random 
neighbours (the average
number in the network is $K = \avg{K_i}$) coupled
by probabilistic synapses $P_{ij} \in [0,1]$ 
(here we use $K=10$, for implementation 
details see~\cite{Kinouchi2006,Costa2015,Campos2017}).
If the presynaptic neuron $j$ fires at time $t$, 
with probability $P_{ij}$ the postsynaptic neuron $i$
fires  at $t+1$. The update is done in parallel, so the 
synapses are multiplicative, not additive like 
in the stochastic neuron model. 
All neurons that fire go to silence
in the next step. Only neurons
in the silent state can be induced to fire. 
Since $1- P_{ij}X_j$ is the probability that neighbour $j$
does not induces the firing of neuron $i$,
we can write the update rule as:
\begin{equation}
P(X_i[t+1] =1 ) = (1-X_i[t]) \left[1 - \prod_j^{K_i} 
(1 - P_{ij} X_j ) \right] \:,
\end{equation}
The control parameter of the static model is the branching ratio
$\sigma = K \avg{P_{ij}}$ and the critical value is $\sigma_c =1$.

In the cellular automata (CA) model with LHG synapses, 
we have~\cite{Costa2015,Campos2017}:
\begin{equation}
P_{ij}[t+1] = P_{ij}[t] + \frac{1}{\tau} \left( \frac{A}{K}
- P_{ij}[t]\right) - u P_{ij}[t] X_j[t] \:. \label{Pij}
\end{equation} 

\subsection{Mean field stability analysis}

The 2d MF map close to the stationary state is:
\begin{eqnarray}
\label{rho_map}
\rho[t+1] &=& (1-\rho[t]) \left[ 1-\left(1 - 
\frac{\sigma[t] \rho[t]}{K} \right)^K \right]\:,\\
\label{sigma_map}
\sigma[t+1] &=& \sigma[t] + \frac{1}{\tau}
\left( A - \sigma[t] \right) 
- u \sigma[t] \rho[t] \:,
\end{eqnarray}
where in the second line we multiplied equation~(\ref{Pij}) by $K$
and averaged over the synapses.

There is a trivial absorbing fixed point $(\rho^0,\sigma^0) = (0,A)$,
stable up to $A=1$, see Supplementary Material.
For $A > 1$, there exists a single stable
fixed point given implicitly by:
\begin{eqnarray}
\label{rho_fixed_point}
\rho^*&=&\left(1-\rho^*\right)\bigg[1-\bigg(1-\frac{A\rho^*}
{(1+u\tau\rho^*)K}\bigg)^K\bigg]\:,\label{sigma_fixed_point}\\
\sigma^*&=&\frac{A}{1+u\tau\rho^*}\:, \label{rhoKC}
\end{eqnarray}
which we can find numerically. 

For this model we have numerical but no analytic results.
However, we can find an approximate solution close to
criticality, where $\rho^*$ is small (see Supplementary Information).
Expanding in powers of $1/\tau$, we get:
\begin{equation}
|\lambda^{\pm}| = 1-\left(\frac{(A-1)
(2K-1)}{2uK}+\frac{1}{2}\right)\:\frac{1}{\tau} + O(\tau^{-2}).
 \label{Lambda}
\end{equation}
This confirms that the same scenario of a weakly stable focus
appears in the CA model.

This last result is particularly interesting. If we put $K=N-1$,
the CA network becomes a complete graph. Performing the limit
$N \rightarrow \infty$, equation~(\ref{Lambda}) gives:
\begin{equation}
|\lambda^{\pm}| = 1 -  \frac{A-1+u/2}{u\tau} 
+  O(\tau^{-2}) \:,
\end{equation}
which is identical to the $\left|\lambda^{\pm}\right|$
value for the LHG stochastic neuron model, 
see equation~(\ref{lambLHG}).

Concerning the simulations, we must make a technical observation.
In contrast to the static model~\cite{Kinouchi2006},
the relevant indicator of criticality here is
no longer the branching ratio $\sigma$, but 
the principal eigenvalue $\Lambda \neq \sigma$ of the synaptic 
matrix $P_{ij}$, with $\Lambda_c 
=1$~\cite{Larremore2011,Wang2017}.
This occurs because the synaptic dynamics creates correlations
in the random neighbour network~\cite{Campos2017}.
So, the MF analysis, where correlations are disregarded, 
does not furnish the exact $\sigma$ or 
$\Lambda$ of the CA model. 
But there exists an annealed version of the model in which
$\sigma^* = \Lambda^*$ and the MF analysis fully
holds~\cite{Costa2015,Campos2017}. 

In this annealed version,
when some neuron fires, the depressing term $-u P_{ij}$ is 
applied to $K$ synapses randomly chosen in the network.
This is not biologically realistic, but destroys correlations
and restores the MF character of the model.
Since all our analyses are done at the MF level, we prefer to 
present the simulation results using the annealed model.
Concerning the SO phenomenology, there is no qualitative
difference between the annealed and the original 
model defined by Eq.~(\ref{Pij}).

\begin{figure}[ht]
\begin{center}
\includegraphics[width=\columnwidth]{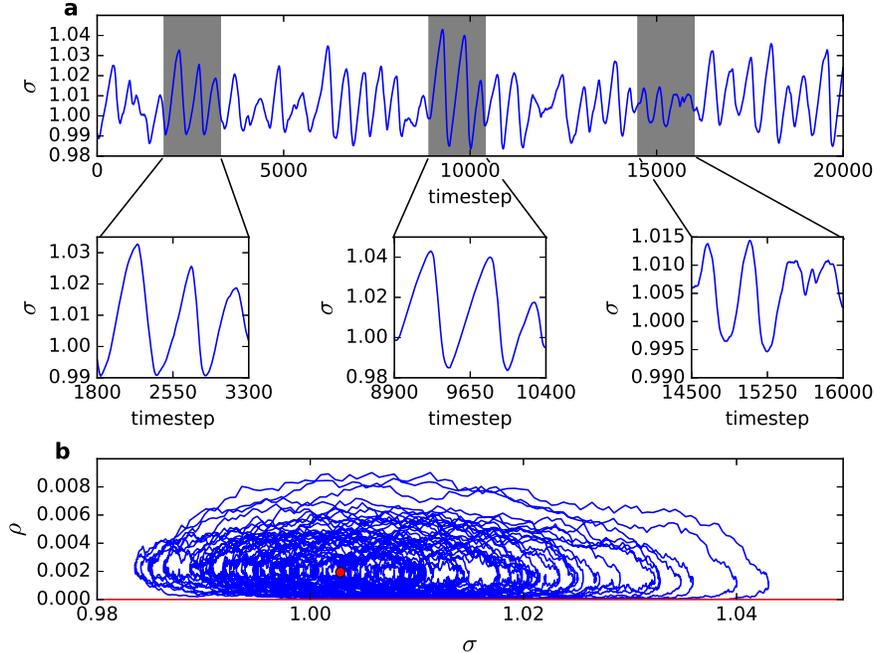}
\caption{{\bf Stochastic oscillations for the 
annealed CA model with LHG synapses.} 
{\bf a}, The SO for $\sigma[t]$ has a sawtooth 
profile where large amplitudes have 
low frequency ($N=128,000$ neurons, $\tau = 500$, 
$A = 1.1$  and $u=0.1$). 
{\bf b}, The SO in the phase plane $\rho$ vs $\sigma$. The 
red bullet is the fixed point given by 
Eqs.~(\ref{sigma_fixed_point}) and˜(\ref{rhoKC}).}
\label{HSO-KC}
\end{center}
\end{figure}

\begin{figure}[ht]
\begin{center}
\includegraphics[width=\columnwidth]{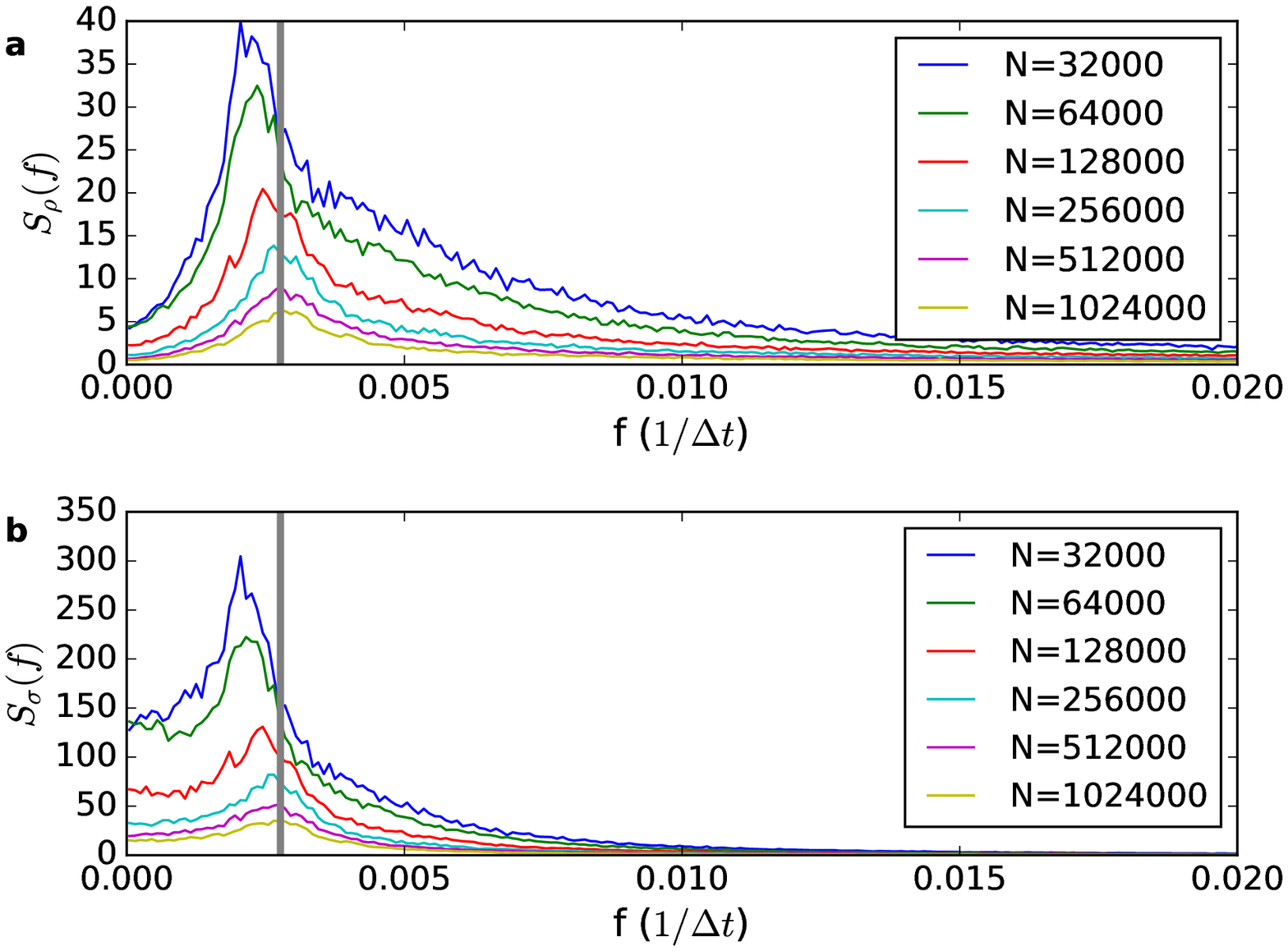}
\caption{{\bf Power spectrum of the time evolution of 
$\rho$ and $\sigma$ in the annealed CA model with
LHG synapses.} 
{\bf a} and {\bf b}, power spectrum of the time evolution 
of $\rho$ and $\sigma$, respectively, for different 
system sizes ($\tau=320$, $A=1.1$, $u = 0.1$). 
The vertical lines mark the theoretical value of the 
frequency of small oscillations of the evolution of the MF 
map near the fixed point calculated from Eq.~(\ref{omegaKC}).}
\label{power_spectrum}
\end{center}
\end{figure}

The full stochastic oscillations in a system 
with $N=128,000$ neurons
and $\tau = 500$ can be seen in Fig.~\ref{HSO-KC}a and b. 
The angular frequency for small amplitude oscillations 
is given by equation~(\ref{omegaKC}) 
(see Supplementary Information).
We have $\omega\simeq\sqrt{(A-1)/\tau}$ for large $\tau$,
which is the same behavior found for stochastic neurons. 
The power spectrum of the time evolution of $\rho$ and $\sigma$ 
is shown in Fig.~\ref{power_spectrum}. 
The peak frequency gets closer to the
theoretical $\omega$ (vertical line)
for larger system sizes because the oscillations have 
smaller amplitudes, going to the small oscillations limit.  

\subsection{Dependence of the stochastic oscillations on system size}

Now we ask if the SO survive in the thermodynamic limit. 
In principle, given that fluctuations
vanish in this limit, and we have
always damped focus for finite $\tau$, the SO
should also disappear when $1/N \rightarrow 0$. 
In contrast, Bonachela \emph{et al.} claimed that, 
in the LHG model, the amplitude of the oscillations basically
does not change with $N$ and is non-zero in the thermodynamic
limit~\cite{Bonachela2010}. Indeed, they proposed that this
feature is a core ingredient of SOqC.

\begin{figure}[ht]
\begin{center}
\includegraphics[width=\columnwidth]{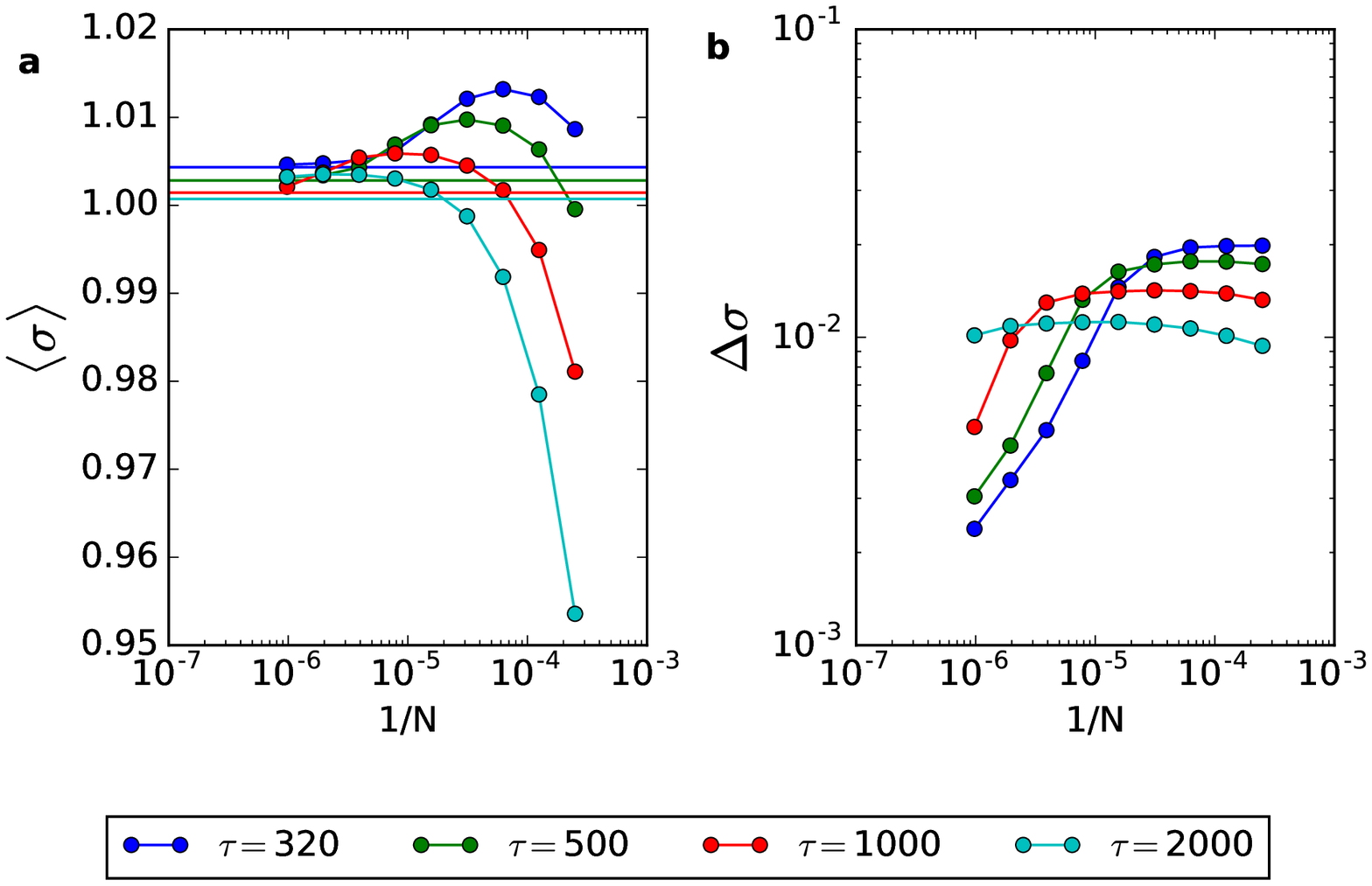}
\caption{{\bf Average and standard deviation of 
the SO time series (CA model) as
a function of $1/N$.} 
{\bf a}, Average $\avg{\sigma[t]}$ for different $\tau$ values. 
Horizontal lines are the value of the fixed points $\sigma^*$
given by equation~(\ref{rhoKC}).
{\bf b}, Standard deviation $\Delta \sigma$ for 
different $\tau$ values. 
Parameters are $A= 1.1$ and $u = 0.1$.
All measures are taken in a window
of $10^6$ time steps after discarding transients.
}
\label{DP-KC}
\end{center}
\end{figure}

In Fig.~\ref{DP-KC}, we measured the average $\avg{\sigma[t]}$ 
and the standard deviation $\Delta \sigma$ 
of the $\sigma[t]$ time series of the annealed model.
We used $\tau = 320, 500, 1,000$ and $2,000$ and system sizes 
from $N = 4,000$ to $1,024,000$. 
We interpret our findings as a $\tau$ dependent crossover
phenomenon due to a trade-off between
the level of fluctuations (which depends on $N$) and the
level of dampening (which depends on $\tau$): 
for a given $\tau$, a small $N$ can produce sufficient
fluctuations so that the SO are sustained without 
change of $\Delta\sigma$. Nonetheless, starting from some $N(\tau)$, 
the fluctuations are not sufficient to compensate
the dampening given by $|\lambda^\pm| < 1$ and
$\Delta \sigma $ starts to decrease for increasing $N$. 
The larger the $\tau$, the less damped is the focus 
and the SO survive without change to
a larger $N$ (the plateau in Fig.~\ref{DP-KC}b 
is more extended to the left).

So, the conclusion is that, for any finite $\tau$, the 
SO do not survive up to the thermodynamic limit, although
they can be observed in very large systems. This is compatible
with Fig.~\ref{power_spectrum} where the power spectrum decreases
with $N$. 
We also see in Fig.~\ref{DP-KC}a that
$\avg{\sigma[t]} \rightarrow \sigma^*$ for increasing $N$, so
the system settles without variance at the MF fixed point 
$(\rho^*,\sigma^*)$ in the infinite size limit. 

We can reconcile our findings with those of the LHG 
model~\cite{Levina2007,Bonachela2010}
remembering an important technical detail: these authors 
used a synaptic dynamics
with $\tau = \tau_0 N$,
in an attempt to have the 
fixed point converging to the critical 
one in the thermodynamic limit
(the same scaling is used in
other models~\cite{Droste2013,Costa2015}).
In systems with this scaling, the 
fluctuations decay with $N$ but, 
at the same time, the dampening controlled by $\lambda(\tau(N))$
also decays with $N$. For this scaling,
we can accord that the SO survive in the thermodynamic 
limit. However, we already emphasized that this
is a non-biological and non-local scaling 
choice~\cite{Campos2017,Brochini2016,Costa2017} because
the recovery time $\tau$ must be finite and the knowledge of the
network size $N$ is non-local.

However, from a biological point of view, this discussion
is not so relevant: although the finite-size fluctuations
(called demographic noise in the 
literature~\cite{McKane2005,Risau2007,Challenger2014,Parra2014})
disappear for large $N$, external (environmental) 
noise of biological origin, not included in the models,
never vanishes in the thermodynamic limit.
So, for practical purposes, the SO and the associated
dragon kings would always be present in 
more realistic noisy networks and experiments. 

\section{Discussion and Conclusion} 

We now must stress what is new in our findings. 
Standard SOC models are related to static systems presenting
an absorbing state phase 
transition~\cite{Dickman1998,Jensen1998,Dickman2000,Pruessner2012}. 
At the MF level, these static systems
are described by an 1d map $\rho[t+1] = F(\rho[t])$
for the density of active sites,
and the phase transition corresponds to a 
transcritical bifurcation
where the critical point is an indifferent node.
This indifferent equilibrium enables the occurrence 
of scale-invariant fluctuations in the $\rho[t]$ variable, 
that is, scale-invariant avalanches but not stochastic
oscillations.

In aSOC networks, the original control parameter of the static system 
turns out an activity dependent variable leading to 2d MF maps.
We performed a MF fixed point stability analysis for three systems: 1)
a stochastic neuronal model with one-parameter 
neuronal gain dynamics, 2)
the same model with LHG gain dynamics, and 3) the CA model with
LHG synapses. For the first one we obtained very simple and 
transparent analytic results; for the LHG dynamics we also
got analytic, although more complex, results; finally, for the
CA model, we were able to obtain a first order approximation
for large recovery time $\tau$. 
Curiously, in the limit of $K = (N-1) \rightarrow 
\infty $ neighbours, the stochastic neuron model and the 
CA model have exactly the same first order leading term. 
The complex eigenvalues have modulus
$|\lambda^\pm| \approx 1 - O(1/\tau)$ and, for large $\tau$,
the fixed point is a focus at the border 
of indifference. This means that, in finite size systems,
the dampening is very low and 
fluctuations can excite and sustain stochastic 
oscillations.

About the generality of our results, we conjecture
they are generic and valid for the whole
class of aSOC models, from networks with discrete deletion-recovery 
of links~\cite{Meisel2009,Meisel2012,Droste2013} 
to continuous depressing-recovering synapses~\cite{Levina2007,Bonachela2010,Costa2015,Campos2017} and 
neurons with firing rate 
adaptation~\cite{Brochini2016,Costa2017,Costa2018}). 
At the mean-field level, all of them are described by
similar two-dimensional dynamical systems: one variable for the order parameter, another for the adaptive mechanism.
For example, the prototypical LHG model~\cite{Levina2007,Bonachela2010} 
uses continuous time LIF neurons (which is equivalent to 
setting $\Gamma \rightarrow \infty$ and $\theta, \mu > 0$ in our model). 
Although with more involved calculations, in principle
one could do a similar mean-field calculation and obtain
a 2d dynamical system with an almost unstable focus
close to criticality, explaining the SO observed in that model.
In another aspect of generality,
stochastic oscillations have been observed previously in
other SOqC systems 
(forest-fire models~\cite{Grassberger1991,Vespignani1998} and some 
special sandpile models~\cite{DiSanto2016,Saeedi2018})
that share a similar MF description with aSOC systems.
Indeed, it seems that stochastic oscillations is a distinctive
feature of self-organized quasi-criticality,
as defined by Bonachela {\it et al.}~\cite{Bonachela2009,Bonachela2010}.

We also emphasize that the
fact that our network has only excitatory neurons is not a limitation
of this study. In a future work~\cite{Kinouchi2018b}, 
we will show that our
model can be fully generalized to an excitatory-inhibitory network 
very similar to the Brunel model~\cite{Brunel2000}, 
with the same results.

So, it must be clear that what is new here is not the 
stochastic oscillations (quasicycles) in biosystems, 
since there is a whole literature about 
that~\cite{Nisbet1976,McKane2005,Risau2007,Wallace2011,
Baxendale2011,Challenger2014,Parra2014}.
What is new here is the interaction of the SO with a critical
point with an absorbing state. This interaction interrupts the
oscillations, producing the phenomenology of avalanches and 
dragon kings. This is our novel proposal for a mechanism
that produces dragon kings coexisting with limited 
power law avalanches.

In conclusion, contrasting to standard SOC, aSOC systems
present stochastic oscillations that will not
vanish in the thermodynamic limit if external
(environmental) noise is present. 
But what seems to be a shortcoming
for neural aSOC models could turn out to be an advantage.
Since the adaptive dynamics with large $\tau$
has good biological motivation,
it is possible that SO are experimentally observable, 
providing new physics beyond the standard model for SOC.
The presence of the Neimark-Sacker-like bifurcation
affects the distribution of avalanche sizes
$P(S)$, creates dragon king events, and all this
phenomenology can be measured~\cite{Arcangelis2012,Poil2008,Poil2012}.
In particular, our raster plots results are 
compatible with system-sized events
(synchronized fires) recently observed in 
neuronal cultures~\cite{Orlandi2013,Yaghoubi2018}.
So, we have experimental predictions 
that differ from standard SOC based on a transcritical
bifurcation (and also from criticality models 
with Griffiths phases~\cite{Moretti2013,Girardi2016}).
We propose that the experimental detection of stochastic 
oscillations and dragon-kings should be the 
next experimental challenge in the field of neuronal 
avalanches.

Finally, we speculate that non-mean field models (square or
cubic lattices) with dynamic gains and SO could be applied to
the modeling of dragon-kings 
(large quasiperiodic "characteristic") earthquakes
that coexist with power-law Gutemberg-Richter distributions 
for the small events~\cite{Carlson1991,Wesnousky1994}.
This will be pursued in another work.

\section*{Methods}

{\bf Numerical Calculations:} 
Numerical calculations were done by using
MATLAB softwares.
{\bf Simulation procedures:} Simulation codes were 
made in Fortran90 and C++11.

\bibliographystyle{plain}
\bibliography{BibSOinaSOC}

\section*{Acknowledgements}
This article was produced as part of the activities of
FAPESP Research, Innovation and Dissemination Center
for Neuromathematics (Grant No. 2013/07699-0, S. Paulo
Research Foundation). We acknowledge financial support
from CAPES, CNPq, FACEPE, and Center for Natural and
Artificial Information Processing Systems (CNAIPS)-USP.
LB thanks FAPESP (Grant No. 2016/24676-1). AAC thanks FAPESP
(Grants No. 2016/00430-3 and No. 2016/20945-8).

\section*{Author contributions}
AAC and JGFC performed the simulations and prepared all
the figures. JGFC, LB, MC and OK made the analytic calculations.
All authors contributed to the writing of the manuscript.

\section*{Competing interests}
The authors declare that they have no competing financial interests.

\section*{Supplementary Information}

\subsection*{Stability analysis of the stochastic 
neuron model with one parameter ($\tau$) gain dynamics}

The MF map is:
\begin{eqnarray}
\rho[t+1] &=& \frac{\Gamma[t] W \rho[t] (1 - \rho[t])} 
{1+\Gamma[t] W \rho[t]} = F(\rho[t],\Gamma[t])\:, \\
\Gamma[t+1] &=& \left(1+ \frac{1}{\tau}
- \rho[t] \right) \Gamma[t] = G(\rho[t],\Gamma[t]) \:. \label{SIG}
\end{eqnarray}
Notice that $(\rho^0 = 0, \Gamma)$ is not a fixed point.
The single fixed point is:
\begin{equation}
\rho^*  =  \frac{1}{\tau}\:,\:\:\:\:\:\:\:\:\:\:\:\:\:\:\:\:\:
\Gamma^* =  \frac{\Gamma_c}{1-2/\tau} \: .
\end{equation}

To perform the fixed point stability analysis 
we calculate partial derivatives  evaluated at the 
fixed point (denoted by $\left.\right]_*$):
\begin{eqnarray}
a=\left.\frac{\partial F}{\partial \rho}\right]_* &=&
\frac{\Gamma^* W - 2 \Gamma^* W \rho^* - (\Gamma^* W \rho^*)^2}{\left[1 + \Gamma^* W \rho^*\right]^2} = 
\frac{\tau-3}{\tau-1} \:, \\
b=\left.\frac{\partial F}{\partial \Gamma}\right]_* &=& 
\frac{W \rho^* (1 - \rho^*)}{\left[ 1 + \Gamma^* W \rho^* \right]^2}  =  \frac{W (\tau-2)^2}{\tau^2 (\tau-1)} \:,\\
c=\left.\frac{\partial G}{\partial \rho}\right]_* &=&
- \Gamma^* = - \frac{\tau}{W(\tau-2)}  \:,\\
d=\left.\frac{\partial G}{\partial \Gamma}\right]_* &=& 
1 + 1/\tau - \rho^* = 1\:.
\end{eqnarray}
The eigenvalues of the system Jacobian matrix 
at the fixed point $(\rho^*,\Gamma^*)$ are determined by
\begin{equation}
 \det(J-\lambda I)=(a-\lambda)(d-\lambda)-b c=0 \label{det} \:.
\end{equation}
So, the characteristic equation is:
\begin{equation}
\lambda^2 -\left[ \left(\frac{\tau-3}{\tau -1} \right)+
1 \right]\lambda + \frac{\tau-3}{\tau-1} + 
\frac{\tau-2}{\tau(\tau-1)} = 0  \:.
\end{equation}
That can be written as:
\begin{equation}
\tau (\tau-1) \lambda^2 - (2 \tau^2 -4\tau) 
\lambda + \tau^2-2\tau -2 =0\: , 
\end{equation}
with solutions:
\begin{eqnarray}
\lambda^\pm &=& \frac{\tau^2-2\tau \pm 
\sqrt{\Delta}}{\tau (\tau-1)} \:,\\
\Delta &=& (\tau^2-2\tau)^2 - \tau(\tau-1)(\tau^2-2\tau-2)
\end{eqnarray}
This equation has no meaning for $\tau < 2$ since this would
give $\rho^* = 1/\tau > 1/2$. When $2 < \tau < 2+\sqrt{2} $ we have $\Delta > 0$, 
$\lambda$ real and $|\lambda | < 1$, that is, the fixed point 
$(\rho^*, \Gamma^*)$ is a stable node.
However, when $\tau > 2+ \sqrt{2} = 3.4142\ldots$, we have
$\Delta < 1$ and two complex
roots $\lambda^\pm$ with $| \lambda | = 
\sqrt{\lambda^+\lambda^-} < 1 $, that is, we have a stable focus with:
\begin{eqnarray}
\lambda^\pm &=& \frac{\tau^2-2\tau \pm i 
\sqrt{-\Delta}}{\tau(\tau-1)}\:,\\
|\lambda^\pm|&=& \sqrt{\lambda^+\lambda^-} =  
\sqrt{ \frac{\tau^2 - 2\tau -2}{\tau(\tau-1)}  }  =  
\sqrt{1 - \frac{\tau+2}{\tau(\tau-1)}}\:,
\end{eqnarray}
which leads to equation~(\ref{lambsimple}).

For small amplitude (harmonic) oscillations, 
the frequency is given by the angle of the eigenvalues written in polar coordinates. We can write it in terms of the determinant $D$ and trace $T$ of the Jacobian matrix $J$ as 
\begin{eqnarray}
\omega &\equiv& |\omega^\pm| =\arctan{\sqrt{\frac{4 D}{T^2}-1}} =  \arctan \frac{\sqrt{\tau + 2/\tau - 4}}{\tau - 2}
\end{eqnarray}
When $\tau \rightarrow \infty$, 
\begin{eqnarray}
\omega \simeq \sqrt{\frac{1}{\tau}} \label{SimpleOmega}
\end{eqnarray}

\subsection*{Stability analysis of the stochastic neuron model
with LHG gain dynamics}

The MF map is given by:
\begin{eqnarray}
\rho[t+1] &=&  \frac{\Gamma[t] \rho[t] 
( 1- \rho[t])}{1+\Gamma[t] \rho[t]} = F(\rho[t],\Gamma[t]) \:,\\
\Gamma[t+1] & = &  \Gamma[t] 
+ \frac{1}{\tau} \left( A - \Gamma[t] \right) 
- u \Gamma[t] \rho[t] =  G(\rho[t],\Gamma[t])  \:.
\end{eqnarray}

We have two fixed points. The absorbing state fixed point:
\begin{eqnarray}
\rho^0 = 0 \:\:\:\:\:\:\:\:\:\:\:\:\:\:\:\:\:\:\:\:\:\:\:
\Gamma^0 = A\:,
\end{eqnarray}
and the non trivial fixed point $(\rho^*, \Gamma^*)$:
\begin{equation}
\rho^*  =  \frac{A-1}{2 A+\tau u}\:,\:\:\:\:\:\:\:\:\:\:\:\:\:\:
\Gamma^*  =  \frac{2 A+\tau u}{2+ \tau u} \:.
\end{equation}

Partial derivatives of $F$ and $G$ with respect 
to $\rho$ and $\Gamma$ are:
\begin{eqnarray}
\frac{\partial F(\rho,\Gamma)}{\partial \rho} &=& 
-\frac{\Gamma \left(2 \rho +\rho^{2} \Gamma -1\right)}{(\rho 
\Gamma+1)^2}\:,\\
\frac{\partial F(\rho,\Gamma)}{\partial \Gamma} &=& 
-\frac{(\rho -1) \rho }{(\rho  \Gamma+1)^2}\:,\\
\frac{\partial G(\rho,\Gamma)}{\partial \rho} &=& -\Gamma u\:,\\
\frac{\partial G(\rho,\Gamma)}{\partial \Gamma}&=& 
-\frac{1}{\tau } -u \rho  +1\:.
\end{eqnarray}
Evaluating at the trivial fixed point, we get:
\begin{eqnarray}
a&=&\left.\frac{\partial F}{\partial \rho}\right]_0 
= \Gamma = A \:,\label{FP0}\\
b&=&\left.\frac{\partial F}{\partial \Gamma}\right]_0 = 0\;,
\label{FP1}\\
c&=&\left.\frac{\partial G}{\partial \rho}\right]_0 
= - u \Gamma = -u A \:, \label{FP2}\\
d&=&\left.\frac{\partial G}{\partial \Gamma}\right]_0 
= 1 - \frac{1}{\tau} \label{FP3}\:.
\end{eqnarray}
The eigenvalues are found from 
$(a-\lambda)(d-\lambda)-bc = 0$, or:
\begin{eqnarray}
(A-\lambda)(1-1/\tau - \lambda) &=& 0 \:,\\
\lambda^2 - (A+1-1/\tau)\lambda + ( 1 - 1/\tau)A &=& 0\:,
\end{eqnarray}
with solution:
\begin{eqnarray}
\lambda^\pm &=& \frac{(A+1-1/\tau)\pm\sqrt{\Delta} }
{2}\:,\label{lamb}\\
\Delta &=& (A+1-1/\tau)^2 - 4 (1 - 1/\tau)A = [A - 
(1-1/\tau)]^2\:.
\end{eqnarray}
Notice that $\Delta \geq 0$ so the eigenvalues 
are always real:
the fixed point is an attractive or repulsive node. 
The final solution of Eq.~\ref{lamb} is:
\begin{eqnarray}
\lambda^+ &=& A \:,\\
\lambda^- &=& 1 - 1/\tau \:.
\end{eqnarray}
This means that the fixed point $(\rho^0 = 0,\Gamma^0=A)$ 
is stable for $\lambda^+ = A < 1$. The critical point is $A = 1 $.

This analysis show that we can have a subcritical state $(0,A)$,
an attractive node, if we use $A < 1$. On the other hand, 
if we use $A>1$,
the absorbing state looses its stability and the non trivial fixed
point $(\rho^*, \Gamma^*)$ is the unique stable solution.
For $A=1$, the point $(\rho^0 =0,\Gamma^0 =1)$ 
is an indifferent node.

Evaluating the derivatives at the non trivial fixed point, we get:
\begin{eqnarray}
\left.\frac{\partial F}{\partial \rho}\right]_* 
&=& \frac{3 -A+u\tau  }{A+u\tau + 1}\:,\\
\left.\frac{\partial F}{\partial \Gamma}\right]_* 
&=& \frac{(A-1) 
(2+u\tau)^2}{(A+u\tau+1) (2 A+u\tau)^2}\:,\\
\left.\frac{\partial G}{\partial \rho}\right]_* 
&=& -\frac{u (2 A+u\tau)}{2+ u\tau}\:,\\
\left.\frac{\partial G}{\partial \Gamma}\right]_* 
&=& 1-\frac{1}{\tau }- \frac{u (A-1)}{2 A+u\tau}\:, 
\end{eqnarray}

As before, the eigenvalues of the Jacobian matrix at the fixed point $(\rho^*,\Gamma^*)$ are determined by equation~(\ref{det}).
After some algebra, we get:
\begin{eqnarray}
|\lambda^+\lambda^-| &=& a d-b c \nonumber \\
& = &
\frac{(1-1/\tau - \frac{u (A - 1)}{2 A + u\tau}) 
(3- A + u\tau)}{(A + u\tau + 1)} 
+ \frac{u(2+u\tau) (A - 1)}{(2 A + u\tau) (A + u\tau + 1)} \:, 
\nonumber \\
&=& \left( 1 - \frac{1}{\tau} \right) 
\left(1 - \frac{2(A-1)}{A+ u \tau + 1} \right) + \frac{u (A-1)^2}
{(A + u \tau + 1)(2A + u \tau)} \:.
\end{eqnarray}

For large $\tau$, we have:
\begin{equation}
\abs{\lambda^\pm}=\sqrt{\lambda^+\lambda^-}
=1-\frac{2(A-1)+u}{2 u\tau} 
+ O(\tau^{-2})\:, \label{SIlambLHG}
\end{equation}
which means that, for large $\tau$, the map is very close to a 
Neimark-Sacker-like critical point. 
As before, the focus
can be excited by fluctuations, generating the SO.

The frequency for small amplitude oscillations is $
\omega \equiv |\omega^\pm| =  \arctan{\sqrt{\frac{4D}{T^2}-1}} $,
where $D$ and $T$ are respectively the determinant and trace of the Jacobian matrix.  When $\tau \rightarrow \infty$,
\begin{equation}
\omega \simeq \sqrt{\frac{A-1}{\tau}}  \label{omegaLHG}
\end{equation}
which is similar to the $\tau^{-1/2}$ behavior
of the simple gain dynamics, Eq. (\ref{SimpleOmega}).

\subsection*{Stability analysis of the 
Cellular Automata model with LHG synapses}

The 2d MF map near the stationary state is:
\begin{eqnarray}
\rho[t+1] &=& (1-\rho[t]) \left[ 1-\left(1 - 
\frac{\sigma[t] \rho[t]}{K} \right)^K \right]
= F(\rho[t],\sigma[t])\:,\\
\label{SIsigma_map}
\sigma[t+1] &=& \sigma[t] + \frac{1}{\tau}\left( A - \sigma[t] \right) 
- u \sigma[t] \rho[t]  = G(\rho[t],\sigma[t])\:,
\end{eqnarray}
where in the second line we multiplied equation~(\ref{Pij}) by $K$
and averaged over the synapses.
As before, the stability of the map is given by the eigenvalues 
of the Jacobian matrix $J(\rho,\sigma)$ at the fixed points. 
The matrix elements are:
\begin{eqnarray}
\label{J11}
\left.\frac{\partial F}{\partial \rho}\right] &=& 
\bigg[\bigg(1-\frac{\sigma\rho}{K}\bigg)^K-1\bigg]
+(1-\rho)\sigma\bigg(1-\frac{\sigma\rho}{K}\bigg)^{K-1}\:,\\
\label{J12}
\left.\frac{\partial F}{\partial \sigma}\right] &=& 
(1-\rho)\rho\bigg(1-\frac{\rho\sigma}{K}\bigg)^{K-1} \:,\\
\label{J21}
\left.\frac{\partial G}{\partial \rho}\right] &=& -u\sigma \:,\\
\label{J22}
\left.\frac{\partial G}{\partial \sigma}\right] &=& 
1-\frac{1}{\tau}-u\rho \:.
\end{eqnarray}
Thus, we can plug in the solutions of 
equations~(\ref{rho_fixed_point}) and (\ref{sigma_fixed_point}) 
in equations~(\ref{J11}-\ref{J22}) and find the eigenvalues of 
$J(\rho,\sigma)$.

Similarly to the previous model, the MF map of the CA with 
LHG synapses has two fixed points. 
The first one is the absorbing state fixed point and is given by
\begin{eqnarray}
\rho^0 = 0 \:\:\:\:\:\:\:\:\:\:\:\:\:\:\:\:\:\:\:\:\:\:\:
\sigma^0 = A\:.
\end{eqnarray}
The second one is nontrivial and must be solved numerically. 

We can compute the elements of the Jacobian matrix 
at the absorbing state fixed point
\begin{eqnarray}
\left.\frac{\partial F}{\partial \rho}\right]_0& =& A \:,\\
\left.\frac{\partial F}{\partial \sigma}\right]_0& = & 0\;,\\
\left.\frac{\partial G}{\partial \rho}\right]_0 &
= & -u A \:,\\
\left.\frac{\partial G}{\partial \sigma}\right]_0 &
= & 1 - \frac{1}{\tau} \:,
\end{eqnarray}
which is the same result of Eqs.~(\ref{FP0}--\ref{FP3}).
This matrix has two eigenvalues: $1-1/\tau$ and $A$. 
The eigenvalue $1-\tau$ corresponds to the $\sigma$ axis, 
which is a stable direction. The eigenvalue $A$ corresponds 
to a unstable (stable) direction for $A>1$ $(A<1)$.

If we assume that we are close to criticality, where 
$\rho^*$ is small, we can find an approximate 
solution for $J(\rho,\sigma)$ at the nontrivial fixed point. 
Expanding the binomial in equation~(\ref{rho_map}) to first order 
and solving for $\rho^*$, we find:
\begin{eqnarray}
\rho*&=&\frac{A-1}{A+u\tau}\:,\\  \label{SIrhoKC}
\sigma^*&=&\frac{A+u\tau}{1+u\tau} = 1 + \frac{A -1}{1+u \tau}\:,
\end{eqnarray}
where we used equation~(\ref{sigma_fixed_point}) 
to compute $\sigma^*$.
Note that, as before, this solution exists only if $A-1>0$. 
This is the exact condition to the absorbing state 
solution be unstable. 

Inserting in equations~(\ref{J11}-\ref{J22}), 
we have:
\begin{equation}
\label{jacob}
J=\left(
 {\begin{array}{cc}
   1-a & b \\
   -c & 1-d \\
  \end{array} }
\right),
\end{equation}
where
\begin{eqnarray}
a&=&\frac{(2K-1)(A-1)}{K(1+u\tau)}\:,\\
b&=&\frac{A-1}{(A+u\tau)^2}\bigg(1+u\tau-\frac{(A-1)(K-1)}{K}\bigg)\:,\\
c&=&\frac{u(A+u\tau)}{1+u\tau}\:,\\
d&=&\frac{1}{\tau}+\frac{u(A-1)}{A+u\tau}\:.
\end{eqnarray}
Note that they are all positive quantities. The eigenvalues of $J$ are:
\begin{equation}
\label{lambdaJ}
\lambda^{\pm}=1-\frac{a+d}{2}\pm\frac{1}{2}\sqrt{(a-d)^2-4bc}.
\end{equation}

For systems to be close to criticality, we must have $\tau\gg 1$. 
Thus, $a$, $b$ and $d$ are of $O(1/\tau)$, while $c=O(1)$. 
Therefore, the $4bc$ term dominates the square root argument 
in equation~(\ref{lambdaJ}). 
Hence, we have two complex conjugate eigenvalues with modulus 
and argument given by:
\begin{equation}
\left|\lambda^{\pm}\right| =\sqrt{1-(a+d)+da+bc}\:.
\label{freq}
\end{equation}
We can expand $|\lambda^{\pm}|$ in powers of $1/\tau$:
\begin{equation}
|\lambda^{\pm}| = 1-\left(\frac{(A-1)
(2K-1)}{2uK}+\frac{1}{2}\right)\:\frac{1}{\tau} + O(\tau^{-2}).
\end{equation}
This confirms that the same scenario of weakly stable focus
appears in the CA model. The harmonic frequency is:
\begin{equation}
\omega \equiv |\omega^\pm| =
\arctan{\frac{\sqrt{4bc-(a-d)^2}}{2-(a+d)}}\:.
\label{omegaKC}
\end{equation}
To first order in $\tau$, we get
$\omega\simeq\sqrt{(A-1)/\tau}$,
which shows that the frequency of oscillations is vanishingly small for large $\tau$, being identical to equation~(\ref{omegaLHG}).

\end{document}